\documentstyle[aps,prl,multicol,epsf]{revtex}

\def\ket#1{| #1\rangle}
\def\bra#1{\langle #1 |}

\title{\rightline{ \small DAMTP-2001-27 \normalsize } \rightline{}
\centerline{Implications of Teleportation for Nonlocality} }
\author{Jonathan Barrett}
\address{Centre for Mathematical Sciences, Wilberforce Road, Cambridge
CB3 0WA, United Kingdom}

\begin{document}
\maketitle

\begin{abstract}

Adopting an approach similar to that of Zukowski [Phys. Rev. A {\bf
62}, 032101 (2000)],
we investigate connections between teleportation and nonlocality. We
derive a Bell-type inequality pertaining to the teleportation scenario
and show that it is violated in the case of teleportation using a
perfect singlet. We also investigate teleportation using `Werner states' of the
form $\alpha P_s + (1-\alpha) I/4$, where $P_s$ is the
projector corresponding to a singlet state and $I$ is the identity. We find that our
inequality is violated, implying nonlocality, if $\alpha >
1/\sqrt{2}$. In addition, we extend Werner's local hidden variable
model to simulation of teleportation with the $\alpha=1/2$ Werner state. Thus
teleportation using this state does not involve nonlocality even
though the fidelity achieved is $3/4$, which is greater than the
`classical limit' of $2/3$. Finally, we comment on a result of
Gisin's and offer some philosophical remarks on teleportation and
nonlocality generally.

\end{abstract}
\vskip10pt

\begin{multicols}{2}
\section{introduction}\label{intro}

Quantum `teleportation', first introduced in \cite{tele}, is a quantum
mechanical scheme that allows one participant (Alice) to transmit a
quantum state in her possession to another participant (Bob). In the
original version, it is a spin-$\frac12$ state that is sent and the
only resources required are a shared singlet and the capacity for
Alice to send two classical bits to Bob. In this case, Bob ends up
with a state identical with the one Alice begins with. The state in
Alice's possession is randomized, so there is no contradiction with
the no-cloning theorem \cite{peresbook}. A notable feature of the
scheme is that it works even when Alice has no knowledge of the state
she is sending.

Briefly, the procedure described in \cite{tele} (hereafter referred to
as `the standard scheme') works as follows. Let the state Alice wants
to teleport be $\ket{\chi}$ and the shared state have density matrix $\rho$. Alice performs a joint
measurement on her half of $\rho$ and $\ket{\chi}$. The measurement projects onto the Bell basis. Alice sends
two classical bits to Bob informing him of which of the four possible
outcomes she got. Bob then performs a corresponding unitary
transformation. If $\rho$ is maximally entangled then these
four unitary transformations can be chosen such that the state Bob
ends up with is identical with $\ket{\chi}$.

If $\rho$ is not maximally entangled then typically, Bob will end up
with something that is not identical with $\ket{\chi}$. Suppose that
at the end Bob is in possession of a state whose density matrix is
$M$. The `fidelity' of the teleportation is defined as
$\bra{\chi}M\ket{\chi}$. In general, the fidelity will depend on
$\ket{\chi}$ but we can define an average fidelity by averaging over
all possible values of $\ket{\chi}$. This average is often referred to
as `the' fidelity for a particular teleportation scheme and shared
state.

In Vaidman's view \cite{vaidman}, quantum teleportation is so called because it involves the transfer of
an `object' from one place to another without it ever being located in
the intervening space. He troubles to argue that
`teleportation' is indeed an appropriate name for this process. This
might already lead us to entertain
rather vague notions that teleportation must be intrinsically
connected with another idea, viz., `nonlocality'. In particular,
in the case in which the fidelity is $1$ (corresponding to perfect
teleportation), it seems intuitively obvious that some sort of
`nonlocality' is at work.

The purpose of this study is to investigate this idea in more
detail (other studies have been undertaken in a similar spirit by
Gisin \cite{gisin}, Zukowski \cite{zuk}, Hardy \cite{hardy} and Cerf
et al \cite{cerf}). First, in Sec. \ref{nonlocality}, we define
more precisely what we mean by `nonlocal' - broadly speaking, we
take it to mean `not simulable by local hidden variables'. In the
literature now, nonlocal is often used simply to mean
entangled. We emphasize that it is the relation between
teleportation and nonlocality in our sense that we shall
investigate. In Sec. \ref{hardy}, we discuss results of Hardy's
\cite{hardy} which show that teleportation and nonlocality are
conceptually distinct and which give us an idea of how local hidden
variables might simulate a teleportation procedure. This leaves open
the question of whether teleportation and nonlocality are physically
distinct - this is the question that we turn to in Sec.
\ref{telenonloc}. Here, we adopt an approach similar to that of
Zukowski \cite{zuk} and derive an inequality, the violation of which
shows that perfect teleportation using a singlet implies nonlocality. In Sec. \ref{wernerteleportation} we consider
teleportation using `Werner states' instead of pure singlets. We
extend Werner's original local hidden variable model in order to
simulate a teleportation scenario with local hidden variables. We also
consider how high the teleportation fidelity can be before
teleportation using a Werner state violates our inequality and
therefore implies nonlocality. Finally, in Sec. \ref{gisin}, we
comment on a result of Gisin's that he claims has relevance to
teleportation and nonlocality. We suggest that his interpretation of
his result is slightly misleading but that the result is still
interesting if interpreted differently.

\section{What We Mean by `Nonlocality'}\label{nonlocality}

We would do well at the outset to specify precisely what we mean by
nonlocality. Broadly, we take it to mean nonsimulablility by local hidden
variables (but see the end of this section for a
qualification). Consider a bipartite state $\rho$ acting on ${\cal
H}_A\otimes {\cal H}_B$. The two subsystems are spatially separated,
one being in the possession of an observer Alice and the other being
in the possession of an observer Bob. Alice performs a measurement
$A$ on her subsystem while Bob performs a measurement $B$ on his;
these measurements occur at spacelike separation from one
another. This procedure is repeated with a new system in the state
$\rho$ each time. We refer to this as a `Bell-type experiment'. If
measurement $A$ has an outcome $a_i$ and measurement $B$ has an
outcome $b_j$, then a hidden variable model supposes that the
probability of getting these two outcomes can be given in the form
\begin{equation}\label{hv}
{\rm Pr}(a_i,b_j|A,B,\rho )=\int\!\!d\lambda~\omega^{\rho}(\lambda
)~{\rm Pr}(a_i,b_j|A,B,\lambda ),
\end{equation}
where $\omega^{\rho }(\lambda )$ is some distribution over a space
$\Lambda$ of hidden states $\lambda$. A local hidden variable model
imposes the additional constraint
\begin{equation}\label{lhv}
{\rm Pr}(a_i,b_j|A,B,\lambda )={\rm Pr}(a_i|A,\lambda ) \, {\rm Pr}(b_j|B,\lambda ).
\end{equation}

It was Bell who first discovered that some quantum states are nonlocal
\cite{bell}. He derived an inequality involving the probabilities of
measurement outcomes that must be satisfied by any local hidden
variable model. He then showed that the quantum mechanical predictions
for two spin-$\frac12$ particles in a singlet state violate this
inequality. Different versions of Bell's inequality were derived in
\cite{ch,chsh}. We refer to these as the Clauser-Horne (CH) inequality
and the Clauser-Horne-Shimony-Holt (CHSH)
inequality, respectively.

These results have since been generalized. In \cite{peres}, Peres
considers experimental scenarios in which Alice and Bob each choose
from an arbitrary finite number of possible measurements to perform,
where each measurement has an arbitrary finite number of possible
outcomes. Peres shows how to construct a list of inequalities, the
idea being that if the outcomes can be simulated with a local hidden
variable model then all the inequalities must be satisfied. He gives
both a graphical method for the easy construction of these
inequalities and an algorithm that produces a complete set - complete
meaning that the satisfaction of all the inequalities is sufficient
for the existence of a local hidden variable model for the particular
experimental scenario considered. We refer to these
inequalities as `Bell-type inequalities'.

Both the formalism developed by Peres and that of Eqs. (\ref{hv})
and (\ref{lhv}) above apply only to a scenario in which Alice and Bob
each perform a single measurement on each run of the experiment. In a
more complicated scenario, Alice and Bob might perform a sequence of
measurements. In \cite{popescuhiddennonloc}, Popescu gives examples of
states for which nonlocality can be revealed in this manner even
though local hidden variable models exist for single measurements
that satisfy Eqs. (\ref{hv}) and (\ref{lhv}) - he calls this `hidden
nonlocality' (we discuss this briefly
again in Sec. \ref{wernerstates}). A new formalism would be needed
to explain what is meant by a local hidden variable model here and
what conditions would have to be violated for the nonexistence of
such a model (see ,e.g., \cite{teufel}). We do not discuss this
further. In what follows, we refrain from calling a state `local' unless
it is completely local, i.e., has no hidden nonlocality (and indeed does
not display what Teufel {\it et al.} call `deeply hidden nonlocality' \cite{teufel}). We
may, however, speak of a state having a `local hidden variable model',
meaning only a model that satisfies Eqs. (\ref{hv}) and (\ref{lhv}) - such
a state may still have hidden nonlocality.

Note that in all of this we have been considering only measurements
performed separately on each particle pair. In some cases, nonlocality
may be revealed if Alice and Bob are allowed collective measurements
on several particle pairs at once, even though the state would be
local if this possibility were not allowed. We choose to ignore this
possibility when classifying states as local or nonlocal - one could
argue that, if collective measurements on $n$ copies of a state $\rho$
are needed to reveal nonlocality, then it is the state $\rho^{\otimes
n}$ that is nonlocal rather than $\rho$.

\section{Conceptually vs Physically Distinct}\label{hardy}

In \cite{hardy}, Hardy shows that teleportation and nonlocality are
conceptually distinct. `Conceptually distinct' means that we ought to
be able to imagine a scenario (not necessarily one complying with
known physical laws but one that is at least logically consistent) in which perfect teleportation is realizable while locality
is demonstrably preserved. Hardy constructs just such a scenario in
the form of a toy theory which allows for notions of systems, states,
measurements, a no-cloning theorem and teleportation. It is also
demonstrably local. Note that this is separate from the question of
whether teleportation and nonlocality are physically distinct -
in other words, it still may be the case that teleportation in quantum
mechanics implies nonlocality.

In fact, Hardy gives two reasons why we might think that nonlocality
is implied by quantum teleportation. The second reason he gives is that
entanglement can be teleported in quantum mechanics. In other words if
Alice and Bob share particles $C$ and $D$ in a singlet state, while
Alice has particles $A$ and $B$ in her possession in some entangled
state $\ket{\psi}_{AB}$, then Alice can teleport particle $B$ to Bob
so that at the end of the protocol $A$ and $D$ are in the state
$\ket{\psi}_{AD}$. (This works essentially because teleportation is a
linear operation applied to particle $B$ \cite{tele}.) It is then
possible to obtain a violation of Bell's inequalities by performing
measurements on $A$ and $D$. However Hardy also responds to this
point:
\begin{quotation}
``We cannot necessarily assert on the basis of this fact that
nonlocality plays a role in quantum teleportation. It is possible that
the extra information which establishes the nonlocal correlations is
only transmitted in the process of measuring the quantities in Bell's
inequalities, and is not transmitted in the teleportation process.''
\end{quotation}
In recognition of this sense that the subsequent testing of
entanglement that has been teleported is not part of the
teleportation process itself, we will hereafter ignore the possibility
of teleporting entanglement. We will consider teleportation of a
single state, not entangled with any others, and the question of when
this might imply nonlocality.

In giving his first reason why teleportation might imply nonlocality,
Hardy notes, following Bennett, that:

\begin{quotation}
``the amount of information needed to specify a general qubit is much
greater than the two bits of information which is classically
communicated during quantum teleportation. One might speculate that
when a qubit is teleported, the extra information is being carried by
the nonlocal properties of the entangled state.''
\end{quotation}
As well as making this point, however, Hardy also has a response:
\begin{quotation}
``On the other hand, it is not possible to extract more classical
information from a qubit than the two classical bits communicated
during teleportation and so there must remain questions about the
reality of the quantum information apparently transmitted during
teleportation.''
\end{quotation}

These two quotations, then, provide us with a motivation for a
detailed investigation of the relationship between teleportation and
nonlocality in quantum mechanics. This is undertaken in the next
section. We adopt an approach similar to that of Zukowski
\cite{zuk}. We derive a Bell-type inequality and show that it is violated in the case of teleportation using a
shared singlet and the standard scheme. 

First, however, Hardy's toy
theory is interesting because it provides some insight into how hidden
variables might be able to describe a teleportation process, so it is
worth examining a few of the details. A particle in Hardy's theory exists in one of four states, labelled
$0$,$1$,$2$ and $3$. The states are `hidden' states because there is
no measurement that will determine the state of a particle
unambiguously and we cannot prepare a particle with an unambiguous
state. We write the state of two particles as a vector
$(x_1,x_2)$, where $x_1,x_2 \in \{ 0,1,2,3 \} $. Two particles may be correlated - for example, we can prepare
them in such a way that their state is given by 25\% chance
of $(0,0)$, 25\% chance of $(1,1)$, 25\%
chance of $(2,2)$ and 25\% chance of $(3,3)$.

This is the preparation used to perform teleportation and is clearly
analogous to the maximally entangled quantum state used in the
standard quantum mechanical scheme (except, of course, it has none of
the nonlocal properties possessed by a maximally entangled quantum
state). Suppose Alice and Bob share two particles, called particles
$2$ and $3$, which have been prepared in this way. Suppose that Alice
also has another particle, particle $1$, in an unknown state, which
she wishes to teleport to Bob. Alice can perform a joint measurement
on particles $1$ and $2$ which can be characterized by four possible
outcomes:

\begin{displaymath}
\begin{array}{cc}
B_0=\left(\begin{array}{cc} 0&0 \\ 1&1 \\ 2&2 \\ 3&3 \end{array}
\right) & B_1=\left(\begin{array}{cc} 0&3 \\ 1&0 \\ 2&1 \\ 3&2
\end{array}\right)
\\
\\
B_2=\left(\begin{array}{cc} 0&2 \\ 1&3 \\ 2&0 \\ 3&1
\end{array}\right) & B_3=\left(\begin{array}{cc} 0&1 \\ 1&2 \\ 2&3 \\
3&0 \end{array}\right).
\end{array}
\end{displaymath}
The notation means that if outcome $B_0$ was obtained, for example, then
the initial state of the two particles was either $(0,0)$, $(1,1)$,
$(2,2)$ or $(3,3)$. Any possible initial state of the two particles
leads to a unique measurement outcome. After the measurement, if outcome $B_0$ was
obtained, then the state of the two particles is one of these four
with equal probabilities i.e., measurement disturbs the system (this
leads to a proof of a no-cloning theorem). The measurement described
here is clearly analogous to to the Bell-measurement used in the
quantum mechanical standard scheme.

Alice now uses two classical bits to inform Bob of her measurement
outcome. If $x_3$ is the initial state of particle $3$ and $x_1$ the
initial state of particle $1$, Bob now knows the value of $i = (x_3 -
x_1) \ {\rm mod \ } 4$. Bob can perform an operation on particle $3$ given by
$U_i: x_3 \rightarrow (x_3 - i) \ {\rm mod \ } 4$. These operations are analogous to the
unitary operations of quantum mechanics. Teleportation has now been
successfully completed - the final state of particle $3$ is identical
with the initial state of particle $1$.

We can take a step back and ask: how exactly has this teleportation
been accomplished? No individual particle has been transmitted from
Alice to Bob (apart from in the preparatory stages and those used to
transmit classical bits). Rather,
correlations between hidden states have been used to ensure that the
final state of the particle in Bob's possession is identical with the
intial state of the particle that Alice teleported. Although Alice in
this theory is unable to determine the state of a particle directly,
it turns out that she is able to do a measurement that tells her the
{\it difference} between the states of particles $1$ and $2$. The
correlations between the hidden states ensure that particle $3$ is
initially in the same state as particle $2$ and Bob is then able to
perform an operation on particle $3$ to make up the difference.

It is clear that this model works in the way it does only because
there are only four different states that a particle can be in - this
corresponds to the fact that Alice sends two classical bits. In this
sense it is artificial and it is not miraculous that it works. It is,
however, quite adequate to establish Hardy's claim that teleportation
and nonlocality are conceptually distinct. In subsequent sections we turn
to consider whether, or in what circumstances,
teleportation in quantum mechanics could be effected by making use of
correlations between hidden variables in a similar manner.

\section{Teleportation and Nonlocality}\label{telenonloc}

Consider the standard scheme for quantum mechanical
teleportation. Alice's first action is to perform a joint measurement on her
half of the shared system and the system she wishes to teleport. The
measurement projects onto the Bell basis. There is another way of
looking at this measurement (described in \cite{mor}). Recall that the
most general type of quantum measurement corresponds to a positive
operator valued decomposition of the identity \cite{peresbook}. We
call such a measurement a POV measurement. In the special case that
the measurement corresponds to a projective decomposition of the
identity, we refer to it as a projective measurement. A POV
measurement on some system can always be realized by attaching an
ancilla and performing a projective measurement on the combined
system \cite{peresbook}. Conversely, any joint projective measurement performed on the
system and ancilla can be thought of as a POV measurement
on just the system. This is exactly what is happening in the case of
teleportation - we regard Alice's
half of the shared state as the system and the state being teleported
as the ancilla. If the state being teleported is $\ket{\chi}=\left(\begin{array}{c}a\\b\end{array}\right)$, then Alice
is performing a POV measurement on her half of $\rho$, the shared
state, with elements as follows:

\vskip30pt
	
\centerline{	
\begin{tabular}{lcc}
	POV element & Bell outcome & \ \ Bob's state, \\
	&& \ \ \ if $\rho = $ singlet \\ \\
	$A_0=\frac12\left(\begin{array}{cc}|b|^2&-ab^*\\-a^*b&|a|^2\end{array}\right)$ & $\ket{\psi_-}$ & $\left(\begin{array}{c}-a\\-b\end{array}\right)$ \\ \\
	$A_1=\frac12\left(\begin{array}{cc}|b|^2&ab^*\\a^*b&|a|^2\end{array}\right)$ & $\ket{\psi_+}$ & $\left(\begin{array}{c}-a\\b\end{array}\right)$ \\ \\
	$A_2=\frac12\left(\begin{array}{cc}|a|^2&-a^*b\\-ab^*&|b|^2\end{array}\right)$ & $\ket{\phi_-}$ & $\left(\begin{array}{c}b\\a\end{array}\right)$ \\ \\
	$A_3=\frac12\left(\begin{array}{cc}|a|^2&a^*b\\ab^*&|b|^2\end{array}\right)$ & $\ket{\phi_+}$ & $\left(\begin{array}{c}-b\\a\end{array}\right)$ \\
\end{tabular}
}	
\vskip30pt
The first column shows the POV element, the second the
corresponding outcome if Alice's measurement is regarded as a
projection onto the Bell basis and the third the state that Bob ends
up with (before he does any unitary transformation) assuming that
$\rho$ is a singlet.

Now suppose that the teleportation procedure is repeated many
times but with a modification: Bob does not bother performing unitary
transformations. Instead, he wishes to determine how close the states
he receives are to those being teleported (he can always take into
account the fact that he \emph{would} have performed a unitary
transformation had he waited for the two classical bits). Bob does
this by performing ordinary projective measurements. He performs them
at spacelike separation from Alice's POV (or Bell basis)
measurements. Also suppose that Alice does not know each time which
state she is teleporting. \emph{Someone} knows which states are being
teleported by Alice and she can be called Clare. From the table above,
it is clear that Alice is actually performing a four-element POV
measurement each time and \emph{which} measurement she is performing
is determined by $\ket{\chi}$ (that is, by the values of $a$ and
$b$). We have now reduced the standard teleportation scenario to a
typical experimental scenario for testing locality via Bell
inequalities. On each run of the experiment, Alice performs one of a
selection of incompatible measurements (although it is Clare who makes
the choice for her) and so does Bob. After many runs, Alice and Bob
can get together and, with Clare's help, see how their results are
correlated. The point is that, if the teleportation is being carried
out with high fidelity, then Bob's results will be strongly correlated
with Alice's and we can expect that some Bell-type inequality will be
violated. If no Bell-type inequality is violated then we can say that
the whole procedure could have been simulated with a local hidden
variable model and that no nonlocality is therefore being
exhibited. The task now is to derive
a Bell-type inequality pertaining to the teleportation scenario and
investigate when it might be violated.

In order to look for a Bell-type inequality that may be
violated, we impose some restrictions. Suppose that Alice only ever
teleports one of two possible states, fed to her randomly by
Clare. This means that in the terms of the analysis above, Alice is
always performing one from a choice of two possible POV measurements,
each of which has four possible outcomes. We restrict Bob to a choice
of two possible projective measurements, each of which has two possible
outcomes. If a joint outcome includes both Alice's and Bob's outcomes, then
there are $(4+4)\times (2+2)=32$ possible joint outcomes. As discussed in
Sec. \ref{nonlocality}, Peres shows how to construct Bell-type
inequalities pertaining to this scenario in
\cite{peres}. Unfortunately, there are a very large number of
such inequalities (the number increases very rapidly with the
number of joint outcomes) and it is hard to know
where to start looking if we want to find one that is violated. For
this reason, we impose another restriction. Suppose that, although Alice's measurements each have
four distinct outcomes, we group them into pairs, so that, for
example, $A_0$ or $A_2$ counts as outcome 1 and $A_1$ or $A_3$ counts as
outcome 2. We then have that Alice has a choice of two measurements,
each with two possible outcomes and so does Bob. This means that we
can apply the well known Clauser-Horne inequality \cite{ch} (or something
equivalent to it) directly to this case. 

First, some terminology. Alice performs one of two
measurements, which we label $T$ and $U$. Alice performing measurement
$T$ corresponds to Clare giving her a state
$\ket{\chi}=\left(\begin{array}{c}a\\b\end{array}\right)$ to
teleport. Alice performing measurement $U$, on the other hand,
corresponds to Clare giving her a state
$\ket{\chi^{\prime}}=\left(\begin{array}{c}a^{\prime}\\b^{\prime}\end{array}\right)$
to teleport. Measurement $T$ has two outcomes, labelled $t$ and
$\bar{t}$. Outcome $t$ corresponds to Alice getting $A_0$ or $A_2$ in
her measurement. Outcome $\bar{t}$ corresponds to her getting $A_1$ or
$A_3$. Similarly, measurement $U$ has two outcomes, $u$ and
$\bar{u}$. $u$ corresponds to outcome $A_0$ or $A_3$ while $\bar{u}$
corresponds to $A_1$ or $A_2$. Note how the grouping of four possible
outcomes into pairs is done differently according to which measurement
Alice is performing. It turns out that unless we do it like this we
do not get an inequality that is violated.

Bob's two measurements are denoted $R$ and $S$, with outcomes
$r$,$\bar{r}$,$s$ and $\bar{s}$. $R$ and $S$ can correspond to spin
measurements in different directions, $r$ and $s$ to spin up results
and $\bar{r}$ and $\bar{s}$ to spin down results.

The inequality we use is
	
	\begin{equation}\label{bell}
	0\leq {\rm Pr}(t,\bar{s})+{\rm Pr}(\bar{u},r)+{\rm
	Pr}(u,s)-{\rm Pr}(t,r)\leq1.
	\end{equation}
Here, ${\rm Pr}(t,r)$ represents the probability of Alice obtaining
outcome $t$ and Bob obtaining outcome $r$ in one run (given that Alice
performs measurement $T$ and Bob measurement $R$). This is seen to be
equivalent to the Clauser-Horne inequality by adding and subtracting
${\rm Pr}(t,s)+{\rm Pr}(u,r)$, leaving

	\begin{eqnarray*}
	0 \leq {\rm Pr}(t)+& &{\rm Pr}(r)+{\rm Pr}(u,s)- \\
	   & & \! {\rm Pr}(t,s)-{\rm Pr}(u,r)-{\rm Pr}(t,r) \leq 1,
	\end{eqnarray*}
where ${\rm Pr}(t)$ is the probability of Alice getting outcome $t$
	and ${\rm Pr}(r)$
the probability of Bob getting outcome $r$.

Now we substitute some values to show that this inequality can be
violated for teleportation using a singlet. Using the standard rule
for obtaining probabilities in quantum mechanics, we get

	\begin{displaymath}
	{\rm Pr}(t,r)={\rm Tr}~[~\rho~(A_0+A_2)\!\otimes\!P~],
	\end{displaymath}
where $P$ is the projection operator corresponding to outcome $r$. We
get similar expressions for the other outcomes. If $\rho$ is a
singlet, this gives

	\begin{equation}\label{ineq}
	0\leq \frac14(2-c(r_x+s_x)+d(r_y-s_y))\leq1,
	\end{equation}
where $c=ab^*+a^*b$ and
$d=-i(a^{\prime}b^{\prime*}-a^{\prime*}b^{\prime})$. We also
have that
$(r_x,r_y,r_z)=\vec{r}$ is the direction of spin measurement $R$ and
$(s_x,s_y,s_z)=\vec{s}$ is the direction of spin measurement $S$. If we set

	\begin{displaymath}
	\left(\begin{array}{c}a\\b\end{array}\right)=\frac{1}{\sqrt{2}}\left(\begin{array}{c}1\\1\end{array}\right),\;
	\left(\begin{array}{c}a^{\prime}\\b^{\prime}\end{array}\right)=\frac{1}{\sqrt{2}}\left(\begin{array}{c}1\\i\end{array}\right),           
	\end{displaymath}
	\begin{displaymath}
	\vec{r}=\left(\frac{1}{\sqrt{2}},\frac{1}{\sqrt{2}},0\right),\;
	\vec{s}=\left(\frac{1}{\sqrt{2}},\frac{-1}{\sqrt{2}},0\right),
	\end{displaymath}	
then Eq. (\ref{ineq}), and therefore Eq. (\ref{bell}), is violated (we get $\frac14(2-c(r_x+s_x)+d(r_y-s_y))=(1-\sqrt{2})/2\approx-0.21<0$). This corresponds to Alice
teleporting one of two states - either spin-up along the x-axis or
spin-up along the y-axis. Bob is performing one of two possible spin
measurements, oriented in the xy-plane at $\pm45^0$ to the x-axis.

Thus we see that perfect teleportation using the standard scheme does
imply nonlocality. In the next section, we introduce `Werner states' and
in Sec. \ref{wernerteleportation}, we investigate when teleportation using a
Werner state might imply nonlocality.

\section{Werner states}\label{wernerstates}

The `Werner states' were first introduced by Werner in
\cite{werner}. They are states in ${\cal H}_d\otimes {\cal H}_d$,
where ${\cal H}_d$ is a
$d$-dimensional Hilbert space, which consist of a mixture of an
entangled pure state in $d\times d$ dimensions and rotationally
symmetric noise represented by the identity. The $d$-dimensional
Werner state is given by
\[
W_d=\frac{1}{d^3}I+\frac{2}{d^2}P^{anti},
\]
where $P^{anti}$ projects onto the totally antisymmetric subspace of
the product space and $I$ is the identity acting on the product space. In his original paper, Werner writes this as
\[
W_d=\frac{1+d}{d^3}I-\frac{1}{d^3}V,
\]
where $V$ is the operator which, acting on any two particle product
state $\ket{\phi}\otimes \ket{\psi}$, `swaps' it to give
$\ket{\psi}\otimes \ket{\phi}$. (Note that $P^{anti}=(1-V)/2$.)

In particular, the two-dimensional Werner state is given by
\[
W_2=\frac18I+\frac12P^s,
\]
where $P^s$ is the projector onto the singlet state.

In \cite{werner}, Werner introduces a local hidden variable model for
these states. We assume that Alice
and Bob share a supply of $d$-dimensional Werner states and that both perform a single projective measurement on each run of a
Bell-type experiment. Briefly, the model works as
follows. Any given pair of particles is assigned a hidden state
$\ket{\lambda }$. $\ket{\lambda }$ is a vector of unit magnitude in
$d$-dimensional complex space. We define a measure $\omega(\lambda )\,
d\lambda $ over the possible
values of $\ket{\lambda }$ which is unchanged under $U(d)$
rotations. This gives us the distribution of $\ket{\lambda }$ values
obtained over many runs of the experiment. Suppose Alice performs a
measurement $A$ with possible outcomes $a_k$ and Bob a measurement $B$
with possible outcomes $b_l$. (We assume that both Alice's and Bob's
measurements are maximal projective measurements - see \cite{mermin}
for how to extend the model, fairly trivially, to nonmaximal projective measurements.) We have that the probability of
obtaining the particular outcomes $a_i$ and $b_j$ is
\[
{\rm Pr}(a_i,b_j|A,B)=\int\!\!d\lambda~\omega(\lambda )~{\rm Pr}(a_i|A,\lambda
)~{\rm Pr}(b_j|B,\lambda ).
\]
We also have:
\begin{eqnarray}\label{rule1}
{\rm Pr}(a_i|A,\lambda)&=&1 {\rm\ if\ }
\bra{\lambda}P_{a_i}\ket{\lambda}<\bra{\lambda}P_{a_{i'}}\ket{\lambda}\quad\forall
i'\not= i \\ &=&0 {\rm\ otherwise} \nonumber,
\end{eqnarray}
and
\begin{equation}\label{rule2}
{\rm Pr}(b_j|B,\lambda )=\bra{\lambda }P_{b_j}\ket{\lambda },
\end{equation}
where $P_{a_k}$ and $P_{b_l}$ are the projection operators
corresponding to measurement outcomes $a_k$ and $b_l$.
It is then reasonably easy to show that
${\rm Pr}(a_i,b_j|A,B)={\rm Tr}[\, W_d\, P_{a_i}\!\otimes\!P_{b_j}]$, which is the
required quantum probability (see both Werner's original paper
\cite{werner} and Mermin \cite{mermin}).

It is significant that the Werner states are mixed, entangled
states. It might seem surprising
that entangled states can have a local hidden variable model,
especially in the light of a result that says that all entangled pure
states are nonlocal \cite{gp,pr}. The story became more complicated
when Popescu showed that the $d$-dimensional Werner states exhibit the
`hidden nonlocality' we spoke of earlier, provided $d\geq 5$ \cite{popescuhiddennonloc,mermin}. By this we mean that
for these states, nonlocality can be revealed if Alice and Bob each
perform a sequence of two projective measurements on each run of the
experiment - note that the local hidden variable model above only
predicts the results for single projective measurements. (More precisely, the only way in which a local theory could
account for the results of the sequential measurments would be for the outcome of
Alice's first measurement to be causally related to the choice of
which measurement to perform for her second measurement - presumably
this is something that a sensible theory should disallow if Alice is a
free observer.) Whether an
alternative scheme might show that the Werner states with $d\leq 4$
have hidden nonlocality or whether the use of POV measurements might
show these states to be nonlocal are open questions.

Rather than consider these questions here, we wish to restrict
attention to the two-dimensional case but generalize slightly so that we
consider states of the form

\[
W_2^{\alpha}=\frac{1- \alpha }{4} I + \alpha \, P^s.
\]
This corresponds to Werner's original state in two-dimensions when
$\alpha =1/2$. It is entangled if $\alpha > 1/3$. Note that it
follows trivially from Werner's local hidden variable model for the
$\alpha =1/2$ case that there also exists a local hidden variable
model (for single projective measurements) for any $\alpha <
1/2$. Rather than consider in full generality which of these
states may be local or nonlocal, in the next section we consider what
teleportation using Werner states might tell us about nonlocality.

\section{Werner States and Teleportation}\label{wernerteleportation}

\subsection{Local Hidden Variable Models}\label{lhvs}

It has sometimes been argued that the state $W_2^{\alpha =1/2}$,
to which Werner's original local hidden variable model applies, must be nonlocal in some
sense because, if used for teleportation, the fidelity achieved is
$\frac34$ which is greater than the `classical limit' of $\frac23$
(see, e.g., Popescu\cite{popescuwhatisnonloc}). This `classical limit'
is obtained as follows. It is in fact the best fidelity that can
be obtained when Alice and Bob do not share any entangled quantum
state and Alice does not know what state she is teleporting. Fidelity
$\frac23$ is achieved by Alice performing a measurement of spin along
the z-axis on her particle and communicating the result to Bob
classically. Bob then prepares a state that is correspondingly spin
up or down along the z-axis. From here on, we refer to teleportation
that works when Alice is ignorant of the state she is teleporting as
`unknown-state teleportation' (this includes, for example, the
standard scheme) and teleportation that requires Alice
to know the state she is teleporting as `known-state
teleportation'. We can say, then, that unknown-state teleportation
with fidelity $>\frac23$ requires Alice and Bob to share an entangled
quantum state. Popescu \cite{popescuwhatisnonloc} describes any instance of unknown-state
teleportation with fidelity $>\frac23$ as a type of
`nonlocality' (presumably in view of the fact that shared entanglement
is required). He is well aware, however, that this would not necessarily imply
violation of a Bell-type inequality and notes therefore that there are
two types of `nonlocality' which are ``inequivalent''. In the context
of our investigation we prefer to
reserve the term `nonlocality' strictly for nonsimulability by a
local hidden variable model, as explained in Sec. \ref{nonlocality}. We can say simply that unknown-state teleportation with
fidelity $>\frac23$ demonstrates that entanglement is present. The
question of whether the $W_2^{\alpha = 1/2}$ state might have
hidden nonlocality or nonlocality that can be revealed by POV
measurements remains open - what we wish to argue here is that its
ability to teleport with fidelity $\frac34$ does not bear on this question.

We do this via a simple modification of Werner's hidden variable
model. Consider the `teleportation'
scenario above in which Bob does not bother waiting for Alice's two
classical bits or performing a unitary transformation but instead
performs a measurement of some kind. If we can provide a local hidden
variable model for this scenario, then we have effectively described
teleportation of fidelity $\frac34$ using local hidden variables (even
though $\frac34$ is greater than the `classical limit').

Werner's original model does not immediately apply because it applies
to projective measurements on the Werner state only. Here, Alice is
performing a Bell-basis measurement on her half of the shared Werner
state and the state to be teleported. As described above, however, for
our purposes we
can regard this as a POV measurement on Alice's half of the shared
state alone. Our aim is to adapt Werner's model so that we recover the
correct quantum mechanical predictions for POV measurements performed
on Alice's half of the shared state at spacelike separation from
projective measurements performed on Bob's half.
We can do this as
follows. Suppose that Alice performs an arbitrary POV measurement $A$,
with elements $A_m$ such that $\sum_m A_m = I$. Then, for a particular
outcome $A_n$, we can define
\begin{equation}\label{modifiedrule1}
{\rm Pr}(A_n|A,\lambda )=\bra{\lambda }A_n\ket{\lambda },
\end{equation}
where on the right-hand side (RHS), $A_n$ represents a positive operator and on the LHS
the corresponding experimental outcome.

If we assume that Werner's original model works, it is easy to show
that this modified model works for an arbitrary POV measurement
performed by Alice. If a spectral decomposition of $A_n$ is given by
$A_n=\sum_i q_i^n Q_i^n$, where $0\leq q_i^n\leq 1$ and $Q_i^n$ is
a projector, then
\begin{eqnarray*}
{\rm Pr}(A_n,b_j|A,B)&=&\int\!\!d\lambda~\omega(\lambda )~{\rm Pr}(A_n|A,\lambda
)~{\rm Pr}(b_j|B,\lambda ) \\
&=&\int\!\!d\lambda~\omega(\lambda )~\bra{\lambda }A_n\ket{\lambda
}~{\rm Pr}(b_j|B,\lambda ) \\
&=&\sum_i q_i^n \!\!\int\!\!d\lambda~\omega(\lambda )~\bra{\lambda
}Q_i^n\ket{\lambda }~{\rm Pr}(b_j|B,\lambda ) \\
&=&\sum_i q_i^n~{\rm Tr}\, [\, W_2^{\alpha =1/2 }\,
Q_i^n\!\otimes\!P_{b_j}] \\
&=& {\rm Tr}\, [\, W_2^{\alpha =1/2}\, A_n\!\otimes\!P_{b_j}],
\end{eqnarray*}
which is the required result. We got from the third line to the fourth
line by assuming that Werner's original model for projective
measurements works.

Incidently, the model can also easily be extended to include the case where Bob
peforms a POV measurement provided Bob's POV elements
commute pairwise. In this case, there exists a single basis in terms of which
we can write down the spectral decomposition of all of Bob's POV
elements. Suppose $B_{n'}=\sum_j q_j^{n'} Q_j$ where the
$Q_j$ are the projectors onto the elements of this basis (and the
point is that they are independent of $n'$). If a projective
measurement corresponding to a decomposition of the identity $\sum_j
Q_j=I$ is performed, Werner's orignal model gives us the
outcome probabilities via rule (\ref{rule1}). For now, we abuse
notation slightly and write these
outcome probabilities as ${\rm Pr}(Q_j|\lambda )$. We can define the outcome
probabilities for Bob's POV measurement as: 
\begin{equation}\label{modifiedrule2}
{\rm Pr}(B_{n'}|B,\lambda
)=\sum_j \, q_j^{n'} \, {\rm Pr}(Q_j|\lambda ).
\end{equation} 
Reasoning similar to that above for
the extension on Alice's side shows that this will reproduce the
correct quantum mechanical probabilities. 

As mentioned above, to our knowledge, no one has
yet constructed a local hidden variable model for $W_2^{\alpha
=1/2}$ which correctly reproduces the quantum mechanical
probabilities for arbitrary POV measurements on both Alice's and Bob's
sides. What we have here is quite adequate for simulation of our
modified teleportation scenario provided Bob is not allowed arbitrary
POV measurements. It seems unlikely that allowing Bob this freedom
would reveal nonlocality where there was none before, but we cannot
claim to have ruled this out. Thus we conclude that with this slight
qualification, the ability of the $W_2^{\alpha =1/2}$ state to
teleport with fidelity $\frac34$ does not betoken nonlocality.

In the next section we consider Werner states with general
$\alpha$. We use the Bell-type inequality derived
in Sec. \ref{telenonloc} to investigate when imperfect
teleportation using Werner states might imply nonlocality.

\subsection{When does Teleportation Imply Nonlocality?}

In this section we use all the same notation as in Sec.
\ref{telenonloc}. Consider again our Bell-type inequality (\ref{bell}):
\[
0\leq {\rm Pr}(t,\bar{s})+{\rm Pr}(\bar{u},r)+{\rm Pr}(u,s)-{\rm Pr}(t,r)\leq 1.
\]
If Alice's and Bob's shared state is $\rho = \alpha P^s +
[(1-\alpha)/4] I$, then we get
\[
0\leq \frac14(2-\alpha (c(r_x+s_x)+d(r_y-s_y))) \leq1,
\]
again with $c=ab^*+a^*b$ and
$d=-i(a^{\prime}b^{\prime*}-a^{\prime*}b^{\prime})$. If, as before, we
set
\[
	\left(\begin{array}{c}a\\b\end{array}\right)=\frac{1}{\sqrt{2}}\left(\begin{array}{c}1\\1\end{array}\right),\;
	\left(\begin{array}{c}a^{\prime}\\b^{\prime}\end{array}\right)=\frac{1}{\sqrt{2}}\left(\begin{array}{c}1\\i\end{array}\right),           
\]
\[
	\vec{r}=\left(\frac{1}{\sqrt{2}},\frac{1}{\sqrt{2}},0\right),\;
	\vec{s}=\left(\frac{1}{\sqrt{2}},\frac{-1}{\sqrt{2}},0\right),
\]
then we have a violation if
\[
\alpha > \frac{1}{\sqrt{2}}.
\]
Thus {\it teleportation using the standard scheme and a Werner state
with \ } $\alpha>1/\sqrt{2}$ {\it \ implies nonlocality}. This of course
implies that these states are nonlocal.

In fact, it is precisely the Werner states with $\alpha >
1/\sqrt{2}$ that violate the ordinary CHSH inequality
\cite{chsh} for a suitable choice of projective measurements, so we
did not need to consider teleportation or derive this inequality
merely to find out that these states are nonlocal. That they violate
the CHSH inequality can
be shown using a result of the Horodeckis' \cite{horocrit} as follows. Any density matrix operating on ${\cal
H}_2\otimes {\cal H}_2$ can be written as:
\[
\rho = \frac14\left(I\otimes I + V^1_i \sigma_i\otimes I + V^2_i I\otimes
\sigma_i + T_{ij} \sigma_i\otimes \sigma_j\right),
\]
where the $\sigma_i$ are the Pauli $\sigma$ matrices, $V^1_i$ and $V^2_i$
are real three-dimensional vectors, $T_{ij}$ is a real
$3\times3$ matrix and repeated indices are summed over. (Note that there are other conditions which
$\vec{V^1}$, $\vec{V^2}$ and $T$ must satisfy for $\rho$ to be a genuine
density matrix). The Horodeckis show that $\rho$ violates the CHSH
inequality for some choice of measurements for Alice and Bob if and
only if $t_1 + t_2 >1$, where $t_1$ and $t_2$ are the largest two
eigenvalues of $T^TT$.

It is easy to show that, in order to write a generalized 2-dimensional Werner
state in this form, we set
\[
\vec{V^1}=0,\ \vec{V^2}=0 {\rm \ and\ } T_{ij}=-\alpha\,\delta_{ij}
\]
to get
\[
W_2^{\alpha}=\frac14\left(I\otimes I-\alpha\, \delta_{ij}\,
\sigma_i\otimes \sigma_j\right).
\]
This state violates the CHSH inequality if and only if
$\alpha>1/\sqrt{2}$, as claimed.

What we have shown in this section, then, is not just that the Werner
states with $\alpha > 1/\sqrt{2}$ are nonlocal - we already
knew this from \cite{horocrit}. We have shown in addition that
consideration of teleportation using these states can also reveal them
as nonlocal. This can be regarded as in keeping with a conjecture made
by Zukowski who suggests that
\begin{quotation}
``the quantum component of the teleportation process cannot be
described in a local and realistic way as long as the initial [shared]
state...neither admits such models.'' \cite{zuk}
\end{quotation}

To summarize, we have a local hidden variable model to describe
teleportation using any Werner state with $\alpha\leq 1/2$ (with
the slight caveat that we have not included the case in which Bob
performs arbitrary POV measurements on the state that he
receives). We have also shown that teleportation with the standard
scheme and a Werner state with $\alpha > 1/\sqrt{2}$ does
imply nonlocality. States with $1/2 < \alpha \leq
1/\sqrt{2}$ do not violate the CHSH inequality for any choice
of projective measurements but apart from this, questions about their
locality remain open. When used for teleportation, they do not violate
our inequality but might violate some other inequality.

Note that a state with $\alpha $ just greater than
$1/\sqrt{2}$ will teleport with a fidelity just greater than $1/2(1+1/\sqrt{2})\approx0.85$. The
teleportation procedure will involve nonlocality even though this
value for the fidelity is below a bound derived by Gisin, which is
$\sim 0.87$ \cite{gisin}. We discuss this further in
Sec. \ref{gisin}.

Our investigation has been resticted to teleportation using Werner
states and the standard scheme. It might be interesting to try to extend these results and find
something more general of the form: unknown-state teleportation implies nonlocality if the fidelity is $F$ or higher and can
otherwise be described using a local hidden variable model. It may
well be, however, that there is no such result to be found. There
could exist two states $\rho$ and $\rho'$ and two teleportation
schemes, which teleport with
fidelities $F$ and $F'$ such that teleportation with $\rho$ implies
nonlocality while teleportation with $\rho'$ does not even though
$F<F'$.

\section{Gisin's Result}\label{gisin}

Before concluding we would like to comment on a
result of Gisin's which he claims has relevance to teleportation and
nonlocality \cite{gisin}. Gisin derives a value for teleportation
fidelity that is given by
\begin{displaymath}
F=\frac12+\sqrt{\frac32}\frac{\arctan\sqrt{2}}{\pi}\approx0.87.
\end{displaymath}
He describes this as an ``upper bound for the fidelity of quantum
teleportation explainable by local hidden variables''. The value of
$F$ is derived as follows.

First Gisin notes that the shared state, $D$, must be local, and ``hence
useless for teleportation''. Then, ``within the local hidden variable
paradigm, Alice could measure the state $\psi_{Alice}$ in the
classical sense of `measuring': finding out what the state
$\psi_{Alice}$ is.'' Here, $\psi_{Alice}$ is the state that Alice is
teleporting (which we earlier called $\ket{\chi }$). In the light of this, to derive the value for $F$
above, we assume that Alice and Bob share nothing, that Alice knows
the quantum state she is trying to teleport and that Alice sends two
classical bits to Bob. The best they can do is to divide the surface
of the Bloch sphere into four regions. Alice uses the two classical
bits to let Bob know which region the state she is teleporting lies
in. Bob then prepares a state in the centre of this region. The
optimal way of doing this is to inscribe a regular tetrahedron in the
Bloch sphere. The areas of the surface of the sphere subtended
by the faces of the tetrahedron are the four regions used. Calculation
then gives the average fidelity obtained as $F$ above.

We feel that the description of $F$ as an ``upper bound for the
fidelity of quantum teleportation explainable by local hidden
variables'' is slightly misleading. On the assumption that
`explainable' here can be replaced with `simulable', the fidelity
`explainable' by local hidden variables rather depends on what is to
count as a simulation of a quantum teleportation procedure. Under our
and Zukowski's approach, we are happy if a local theory can predict the results of
Alice's Bell measurement and of a spacelike-separated measurement made
by Bob (for a completely different approach which is equally
interesting, see \cite{cerf}). Using our and Zukowski's approach, we
found that the ability of the state $W_2^{\alpha = 1/2}$ to
teleport with fidelity $\frac34$ does not betoken any form of
nonlocality. On the other hand, the fact that under the standard scheme, a state
$W_2^{\alpha }$ with $\alpha$ just greater than $1/\sqrt{2}$
teleports with fidelity just greater than
$1/2(1+1/\sqrt{2})\approx0.85$ does betoken
nonlocality. This is despite the value of $\sim 0.85$ being below
Gisin's bound. It is true that, if Alice knows the state she is trying
to teleport, then she can do better than this using only local means -
this is what Gisin's result shows. If Alice does not know the state
she wants to teleport, however, then the standard teleportation scheme
is the best she can do and this will involve nonlocality. (Note that the
standard scheme is indeed the optimal scheme for unknown-state
teleportation using a Werner state; see, e.g., \cite{optimaltele}).

In addition, there is no reason why teleportation with a fidelity
greater than Gisin's bound should not be simulable by local hidden
variables in some cases. The key here is that hidden variable theories are surely
only required to reproduce probabilities for measurement results. In
other words a hidden variable model need not specify an actual quantum
state to be received by Bob on each run of the experiment - we must
assume that Bob does some kind of measurement and it is only the
outcome of this measurement that must be predicted by the model. So it
is not quite correct to say that a shared state that is local must be
``useless for teleportation''. The local correlations may be useful in
helping Alice and Bob achieve correlated measurement results.

At the least, we feel that Gisin's result is less genuinely to do with
locality than our own or Zukowski's \cite{zuk} or those of Cerf {\it et al.}
\cite{cerf}. This does not mean that Gisin's result is uninteresting,
however. Interpreted as the best fidelity achievable when Alice and
Bob share nothing at the start of the protocol, and Alice knows the
state she is trying to teleport and is limited to the sending of two
classical bits, it is correct. It can be contrasted
with the value of $\frac23$ for the fidelity which is the best Alice and
Bob can do when they share nothing at the start of the protocol and
Alice does not know the state she is trying to teleport. 

Gisin's result is also useful. In
the case that Alice and Bob share a nonmaximally entangled quantum
state, they cannot achieve unit fidelity. Gisin's result shows that if
Alice knows the state she is trying to teleport, is limited to the
sending of two classical bits to Bob and the best fidelity
achievable with a quantum scheme is $<F\approx0.87$, then they may as
well not bother using the shared quantum state. They would do better
to use the purely classical scheme above.

\section{Conclusion}\label{conclusion}

Perfect teleportation (i.e., teleportation with unit fidelity) initially seems
paradoxical because only two classical bits are sent yet Bob ends up
with a quantum system in a state identical with the state of Alice's
input system - and it would take an infinite amount of classical
information to specify precisely a quantum state. It is concluded
(rather vaguely) that some sort of nonlocality must be involved - the
extra `information' must be transmitted nonlocally. Vaidman argues
that, correctly interpreted, quantum teleportation involves the
transfer of an `object' from one place to another without it ever
being located in the intervening space \cite{vaidman}. This also
sounds vaguely paradoxical and might suggest nonlocality (although Vaidman
himself is more concerned to reconcile this view of teleportation with his own
belief in a many-worlds type interpretation of quantum mechanics). Rather than adopt either
of these two viewpoints, we are more inclined to dissolve these
paradoxes (at least partially) by sharing Hardy's
doubts concerning the reality of the `information' apparently
transmitted during teleportation (see the quotations in Sec. \ref{hardy}). We suggest that the paradox is
resolved if we consider a quantum state as being a description of an
ensemble of systems, rather than a single system - Bob can identify
the state and any information contained therein by performing
measurements on the whole ensemble. But to teleport the whole
ensemble, Alice does indeed send an infinite number of classical bits.

Having said this, teleportation might still involve nonlocality. If we define nonlocality to mean nonsimulability by local hidden
variables, then to speak meaningfully of teleportation being local or
nonlocal we must have Bob performing a measurement of some sort on
the state that he receives (because hidden variable models are
required only to reproduce the results of measurements). Bob's measurement
is at spacelike separation from Alice's. We can speak of the
teleportation as being nonlocal if Bob's results are correlated with
Alice's in a way that cannot be simulated with a local model. 

In investigating this, we have
considered perfect teleportation using a singlet, derived an
appropriate Bell-type inequality and shown that it is violated. So
perfect teleportation is nonlocal. We have also considered
teleportation using Werner states, of the form $\rho = \alpha P^s
+[(1-\alpha)/4] I$. Using the same inequality, we found that the
teleportation is nonlocal precisely for those Werner states that
violate the CHSH inequality, i.e., those with $\alpha >
1/\sqrt{2}$. These teleport with fidelity $F >
1/2(1+1/\sqrt{2})\approx 0.85$. We also extended Werner's
local hidden variable model for the $\alpha =1/2$ states to give a
local model describing teleportation using these states (the fidelity
of which is $\frac34$). We concluded that the ability to teleport with
fidelity $\frac34$ does not confer nonlocality in this case, with the
qualification that we have not allowed Bob arbitrary POV
measurements. 

Broadly speaking, the status of the (two-dimensional) Werner states with
respect to locality remains unknown. Teleportation shows that those
with $\alpha > 1/\sqrt{2}$ are nonlocal - but we already knew
this. Teleportation using the $\alpha = 1/2$ state can be
simulated locally - but the state may still have a hidden nonlocality
to be revealed by other means. We do not know whether (unknown-state)
teleportation using the $1/2 < \alpha \leq 1/\sqrt{2}$
states can be simulated locally or not, or whether they might have
nonlocality to be revealed by other means.

Related independent results have been very recently circulated by Clifton and Pope \cite{cliftonpope}.

\vskip5pt
{\it Note added.} An LHV model allowing for POV measurements on Bob's side has recently
been constructed for the $\alpha = 5/12$ state, which teleports with
fidelity 17/24 \cite{barrett}. Thus we can simulate teleportation with
fidelity 17/24 without the qualification that Bob is restricted to
projective measurements.
\vskip5pt

\leftline{\bf Acknowledgments}

I am grateful to Trinity College, Cambridge for support, CERN for
hospitality and to the European grant EQUIP for partial support.
I am also indebted to Adrian Kent for much assistance with this work.

\end{multicols}
\end{document}